# Evidence for Unconventional Superconductivity and Nontrivial Topology in PdTe

Ramakanta Chapai[1], P. V. Sreenivasa Reddy[2], Lingyi Xing[1], David E. Graf[3], Amar B. Karki[1], Tay-Rong Chang[2,4,5], and Rongying Jin[1,6,*]

PdTe is a superconductor with $T_c$ ~4.25 K. Recently, evidence for bulk-nodal and surface-nodeless gap features has been reported in PdTe [Yang *et al*., *Phys. Rev. Lett*. 130, 046402 (2023)]. Here, we investigate the physical properties of PdTe in both the normal and superconducting states via specific heat and magnetic torque measurements and first-principles calculations. Below $T_c$, the electronic specific heat initially decreases in $T^3$ behavior (1.5 K < $T$ < $T_c$) then exponentially decays. Using the two-band model, the superconducting specific heat can be well described with two energy gaps $\Delta_1 = 0.372$ meV and $\Delta_2 = 1.93$ meV. The calculated bulk band structure consists of two electron bands ($\alpha$ and $\beta$) and two hole bands ($\gamma$ and $\eta$) at the Fermi level. Experimental detection of the de Haas-van Alphen (dHvA) oscillations allows us to identify four frequencies ($F_\alpha = 65$ T, $F_\beta = 658$ T, $F_\gamma = 1154$ T, and $F_\eta = 1867$ T for $H // a$), consistent with theoretical predictions. Nontrivial α and β bands are further identified via both calculations and the angle dependence of the dHvA oscillations. Our results suggest that PdTe is a candidate for unconventional superconductivity.

[1]*Department of Physics and Astronomy, Louisiana State University, Baton Rouge, LA 70803, USA*
[2]*Department of Physics, National Cheng Kung University, Tainan 701, Taiwan*
[3]*National High Magnetic Field Laboratory, Tallahassee, FL 32310, USA*
[4]*Center for Quantum Frontiers of Research and Technology (QFort), Tainan 70101, Taiwan*
[5]*Physics Division, National Center for Theoretical Sceinces, Taipei 10617, Taiwan*
[6]*Center for Experimental Nanoscale Physics, Department of Physics and Astronomy, University of South Carolina, Columbia, SC 29208, USA*
[*]*Corresponding author: rjin@mailbox.sc.edu*

**Introduction**

Exploring new topological materials has been a vibrant research in condensed matter physics for the past decade due to their associated intriguing physical properties. In particular, topological superconductors and topological insulators are at the frontier of research owing to their potential for applications in quantum computation and spintronic technologies [1-4]. Topological superconductors that host Majorana fermions can be realized on materials possessing topologically nontrivial bands and having the superconducting ground state [5-10]. In recent years, attempts have been made to realize such a state through various protocols such as applying external pressure on topological systems [11, 12], doping a topological insulator [13-15], or fabricating heterostructures consisting of topological insulators and conventional superconductors [16, 17]. However, superconductivity induced by pressure or chemical doping often suffers from the difficulty of achieving high superconducting volume fraction [15,17], and it is extremely challenging to fabricate an atomically sharp interface between two different materials [16, 17]. An effective venue of realizing topological superconductivity is to identify nontrivial topological bands in superconductors [18-20].

Topological superconductors exhibit unconventional superconducting properties, for example, point or line nodes in the gap structure or mixed order parameters [21-23]. Therefore, exploring a material system with nontrivial band structure and unconventional superconductivity is promising for topological superconductivity [18, 24, 25]. Recently, angle-resolved photoemission spectroscopy (ARPES) shows evidence for bulk-nodal and surface-nodeless gap features in PdTe [26]. In this Article, we take an alternative route to study the unconventional superconductivity of PdTe by specific heat measurements down to 50 mK and its topological properties by analyzing quantum oscillations observed in the magnetic torque supported by theoretical calculations. Below $T_c \sim 4.25$ K [27, 28], the electronic specific heat initially decreases with temperature ($T$) in $T^3$ behavior (1.5 K $< T <$ $T_c$) then exponentially decays at $T <$ 1.5 K indicating unconventional superconductivity in PdTe. The calculated bulk band structure consists of two electron bands ($\alpha$ and $\beta$) and two hole bands ($\gamma$ and $\eta$) at the Fermi level, two of which are topologically nontrivial. Measurements of the magnetic torque under high magnetic fields up to 35 T show clear de Haas-van Alphen (dHvA) oscillations. Detailed analysis of the dHvA oscillations allows us to experimentally identify four frequencies ($F_\alpha = 65$ T, $F_\beta = 658$ T, $F_\gamma = 1154$ T, and $F_\eta = 1867$ T), consistent with theoretical predictions. By constructing the Landau fan diagram for each band, we extract the Berry phase, which is nontrivial for the α and γ bands. On the other hand, the Berry phase for the β band changes from trivail for $H$ //$a$ to nontrivial for $H$//$c$. This suggests that PdTe is a candidate for topological superconductivity.

## Results and discussion

### Electrical resistivity, magnetic susceptibility and specific heat

Figure 1(a) shows the temperature dependence of the electrical resistivity (ρ) of single crystalline PdTe between 2 and 8 K. The resistivity suddenly drops to zero at $T_c$ ~ 4.25 K with the half width of the transition of 0.1 K, indicating a superconducting transition. As shown in the inset of Fig. 1(a), above $T_c$, ρ($T$) increases with increasing temperature with ρ(5 K) ~ 1.43 μΩ cm and ρ(300 K) ~ 61.7 μΩ cm. The low residual resistivity and large residual resistivity ratio (*RRR*) (ρ(300 K)/ρ(5 K) ~ 43) reflect the high quality of our single crystals. Note that this value of *RRR* is much larger than that reported earlier for polycrystalline samples [27, 28].

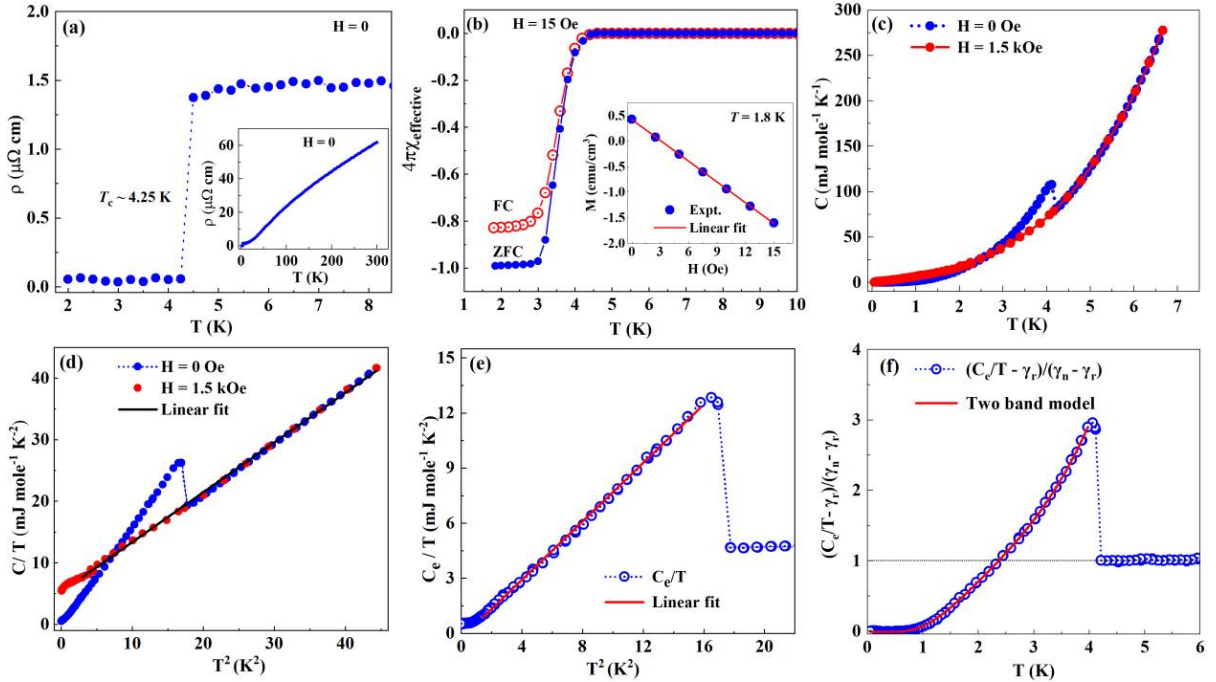

**FIG. 1. Superconductivity in PdTe.** (a) Zero-field electrical resistivity of PdTe at low temperatures showing zero resistance below 4.25 K. Inset: Temperature dependence of the electrical resistivity between 2 and 300 K. (b) Magnetic susceptibility of PdTe in both ZFC and FC modes measured by applying 15 Oe field. Inset: magnetization as a function of $H$ at 1.8 K, where the solid line is a linear fit of $M$ using $M = A(H-H_0)$ for determining the residual field $H_0$. (c) Specific heat (*C*) of PdTe under $H = 0$ and 1.5 kOe. (d) $C/T$ versus $T^2$. The black solid line is the fit of the data under field to the relation $C/T = \gamma + \beta T^2$. (e) Electronic specific heat plotted as $C_e/T$ versus $T^2$. The solid line represents the linear fit of data. (f) ($C_e/T - \gamma_r)/(\gamma_n - \gamma_r)$ versus $T$. The solid line is the fit of data to the two-band model (see text).

To confirm the resistivity drop is due to superconductivity, we measure the magnetic susceptibility. Fig. 1(b) shows the magnetic susceptibility ($\chi$) of PdTe in both the zero-field-cooled (ZFC) and field-cooled (FC) modes. A large diamagnetic signal develops below $T_c$ in both $\chi_{FC}$ and $\chi_{ZFC}$. This confirms that the resistivity drop in Fig. 1(a) corresponds to the superconducting transition. At 1.8 K, the ratio $\chi_{FC}/\chi_{ZFC}$ ~84%, implying a high superconducting volume of the sample. Note that both residual magnetic field and demagnetization factor have been considered with data shown in Fig. 1(b).

In order to uncover the nature of the pairing mechanism in superconducting PdTe, we measure the specific heat down to 0.05 K, two orders lower than $T_c$. Fig. 1(c) shows the temperature dependence of the specific heat ($C$) between 0.05 and 7 K measured in $H = 0$ (blue dots) and $H = 1.5$ kOe (red dots). A sharp jump at $T_c = 4.25$ K indicates the bulk superconductivity. Superconductivity is almost completely suppressed under $H = 1.5$ kOe as seen from the absence of any anomaly down to the lowest temperature measured. We thus fit the $C(T)/T(H = 1.5$ kOe$)$ versus $T^2$ with $C(T)/T = \gamma_n + \beta T^2$ as represented by the black solid line in Fig. 1(d), where $\gamma_n$ is the normal-state electronic contribution and $\beta T^2$ is the lattice contribution to the specific heat. The fitting yields $\gamma_n = 4.77$ mJ mol$^{-1}$K$^{-2}$ and $\beta = 0.81$ mJ mol$^{-1}$K$^{-4}$. Using $\beta = (12/5)NR\pi^4\Theta_D^{-3}$, we can estimate the Debye temperature $\Theta_D$~213 K. This value is similar to our previous estimate [27] but somewhat larger than that reported in Ref. [28]. The lattice contribution is then subtracted from the total specific heat $C(T, H = 0)$ to obtain the electronic $C_e(T, H = 0)$. Figure 1(e) displays the temperature dependence of $C_e/T$ plotted as a function of $T^2$. Between ~1.5 K and $T_c$, $C_e/T$ shows a linear dependence with $T^2$ as represented by the red solid line. For a superconductor with nodes in the energy gap, a power law temperature dependence is expected in $C_e/T$, with the exponent determined by the form of nodes. Point nodes give $C_e/T \propto T^2$, while line nodes give $C_e/T \propto T$ [29]. The $T^3$ dependence of $C_e$ has been considered as evidence for $p$-wave pairing symmetry [30-33].

Below ~1.5 K, $C_e/T$ gradually deviates from the $T^2$ dependence, which can be fitted with the expression $C_e/T = \gamma_r + A^*\exp(-\Delta/k_BT)$. The fitting yields $\gamma_r = 0.65$ mJ mol$^{-1}$K$^{-2}$ and the energy gap $\Delta$ ~ 0.395 meV. The finite residual electronic specific heat $\gamma_r$ can be intrinsic or extrinsic. For a system with 100% superconducting volume, the finite $\gamma_r$ would imply that the pairing gap consists of nodes as discussed above. For PdTe, our magnetic susceptibility data in Fig. 1(b) suggests about 16% non-superconducting volume, which is close to $\gamma_r/\gamma_n$ ~13.6%. Fig. 1(f) plots $C_e(T)$ as $C_{es}/(\gamma_n-\gamma_r)T$ versus $T$ with $C_{es} = C_e-\gamma_r T$. Below $T_c$, $C_{es}/(\gamma_n-\gamma_r)T$ represents normalized superconducting electronic specific heat, which can be well fit by the $T^2$ dependence above 1.5 K (Fig. 1(e)) and exponential $T$ dependence below 1.5 K. At $T_c$, the specific heat jump $\Delta C_{es}/(\gamma_n-\gamma_r)T$ ~ 2.1, much higher than the expected BCS value for a superconductor in the weak-coupling limit [34, 35].

Through the above quantitative analysis of the temperature dependence of $C_{es}$, we find several intriguing features that cannot be explained by the conventional BCS theory. First, the $T^3$ dependence of $C_{es}$ between $T_c/3 < T \leq T_c$ has been seen in unconventional superconductors such as $Sr_2RuO_4$ [33]. While such $T$ dependence is also observed in $HfV_2$ in the similar temperature range, the low-temperature exponential decay of $C_{es}$ suggests that the pairing symmetry is fully gaped with the absence of any nodes and the $T^3$ behavior can be attributed to strong electron-phonon coupling [29, 36]. For PdTe, $C_{es}(T)$ behaves similarly to that of $HfV_2$ and its high $\Delta C_{es}/(\gamma_n-\gamma_r)T$ points to strong electron-phonon coupling nature as well. However, the obtained energy gap gives $\Delta/k_BT_c$ ~ 1.08 for PdTe, even much smaller than the expected BCS predicted value (~1.764) for the weak-coupling limit [34]. Such small $\Delta/k_BT_c$ is usually seen in superconductors with two superconducting energy gaps such as $Ba(Fe_{1-x}Co_x)_2As_2$ [30], $MgB_2$ [37], $NbSe_2$ [38], and FeSe [39]. Fitting data in Fig. 1(f) using the two-band model [30], we obtain $\Delta_1 = 0.372$ meV and $\Delta_2 = 1.934$ meV, or $\Delta_1/k_BT_c = 1.02$ and $\Delta_2/k_BT_c = 5.29$. As demonstrated in Fig. 1(f), the model represented by the red line fits the experimental data very well up to $T_c$.

**Electronic band structure and Fermi surface**

To understand the band topology, we calculate the electronic band structure of PdTe. Fig. 2(a) shows the bulk Brillouin zone (BZ) of PdTe with the relevant high-symmetry points. The calculated bulk band structure along high-symmetry directions is displayed in Fig. 2(b) and the Fermi surfaces projected in the first BZ are shown in Fig. 2(c). The band $\gamma$ is crossing the Fermi level ($E_F$) along Γ-A as highlighted in Fig. 2(b). A bowl-like Fermi surface that displays hole nature around the $A$ point can be seen. The η band consists of two parts: one part is around the $A$ point and the other around Γ with a star shape. Both γ and η band dispersions have the hole nature, while the α and β bands are both around the $K$ point and exhibit electron-like band dispersion. Based on the orbital decomposition analysis (Fig. S5, Supplementary Material), the hole natured Fermi surfaces (γ and η) are mainly dominated by Te $p$-orbital and electronic natured Fermi surfaces (α and β) are dominated with Pd *d*-orbital.

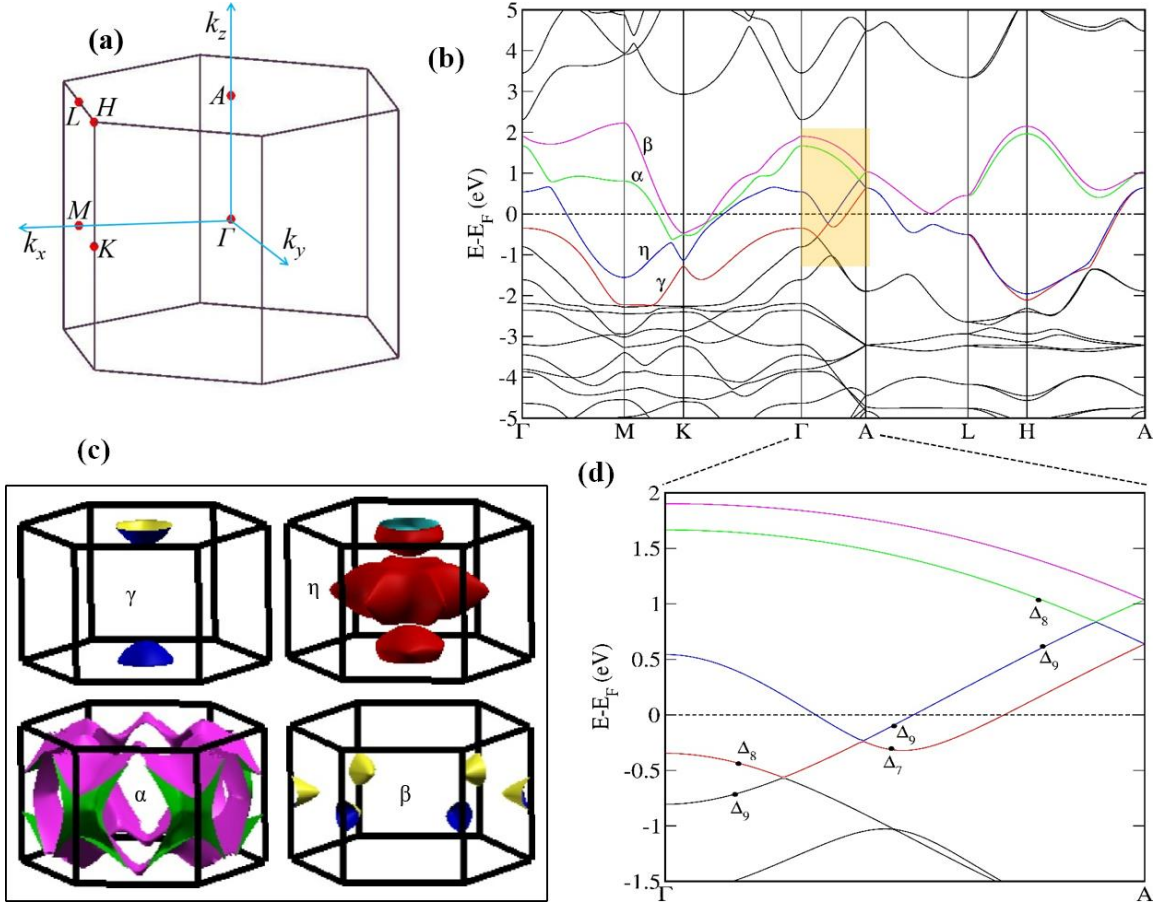

**FIG. 2. Electronic band structure and Fermi surface of PdTe.** (a) Bulk Brillouin zone of PdTe. (b) Calculated bulk band structure of PdTe in the presence of spin-orbit coupling. The bands cross the Fermi level are represented in color. (c) Fermi surfaces of the corresponding band dispersions in (b). (d) An enlarged view of the shaded region in (b). The bands near nodal points are represented with irreducible representations (IRs). At Γ point, the IRs for γ, η, α and β bands are $\Gamma_8$, $\Gamma_7$, $\Gamma_8$ and $\Gamma_9$ respectively. Along Γ-A, these bands transformed to $\Delta_8$, $\Delta_7$, $\Delta_8$ and $\Delta_9$.

As shown in the Fig. 2(b), the conduction and valence bands are separated by a continuous energy gap except at the discrete *k* points near the Fermi level ($E - E_F = 0$), revealing the semimetallic ground state. Figure 2(d) shows the zoomed-in view of the band structure near the crossing points (shaded window in Fig.2(b)). We see that the conduction and valence bands cross each other along the Γ-A direction, forming the Dirac nodes at the middle of the $k_z$ axis. Since PdTe possesses both the time-reversal and space-inversion symmetry, the bulk band structure exhibits doubly degeneracy in the whole BZ. The Γ-A direction is the $C_{3z}$ rotational axis of the bulk BZ. Thus, the rotational eigenvalues of energy states can be well-defined along the Γ-A direction. The space group representations of each branch near the Fermi level are labeled in Fig.

2(d). Specifically, the states of the nearest band structure branches belong to two different representations of the space group. With respect to the $C_{3z}$ rotational axis, the two representations have opposite rotational eigenvalues under the rotational operation. Therefore, gap opening is forbidden at the crossing point between two branches of different rotational eigenvalues, which results in the gapless Dirac nodes. In this sense, the Dirac nodes are under the protection of the $C_{3z}$ rotational symmetry.

**de Haas-van Alphen oscillations**

To confirm the nontrivial topology in PdTe, we carry out magnetic torque measurements at low temperatures and high magnetic fields ($H$). Fig. 3(a) displays the $H$ dependence of the magnetic torque, $\tau(H)$, measured by applying $H$ along the $a$ axis ($H // a$) at 2 K, which exhibits the de Haas-van Alphen (dHvA) oscillations above ~6 T. The emergence of the dHvA oscillations is the consequence of the Landau level formation in the presence of magnetic field [40, 41]. By subtracting the non-oscillatory background, we extract the oscillatory part of the magnetic torque, $\Delta\tau$, as plotted in the inset of Fig. 3(a). Fig. 3(b) shows $\Delta\tau$ plotted as a function of the inverse magnetic field ($H^{-1}$) at 2, 4, 6, 8, and 10 K. The oscillation amplitude decreases with increasing temperature. From the Fast Fourier Transformation (FFT), four oscillation frequencies are identified with $F_\alpha = 65$ T, $F_\beta = 658$ T, $F_\gamma = 1154$ T and $F_\eta = 1867$ T, as shown in Fig. 3(c). The existence of multiple frequencies in the dHvA oscillations is consistent with the multiband nature of PdTe obtained from our DFT calculations (see Fig. 2(b)).

The dHvA oscillations can be described by the Lifshitz-Kosevich (LK) formula [40, 41]

$$\Delta\tau \propto -H^\lambda R_T R_D R_S \sin\left[2\pi\left\{\frac{F}{H} - \left(\frac{1}{2} - \Phi\right)\right\}\right], \qquad (1)$$

where, $F$ is the frequency of an oscillation, $R_T = \frac{A\left(\frac{m^*}{m_0}\right)T}{Sinh(A\left(\frac{m^*}{m_0}\right)T)}$ is the thermal damping factor ($A = \frac{2\pi^2 k_B m_0}{e\hbar H}$) and $R_D = \exp\{-A\left(\frac{m^*}{m_0}\right)T_D\}$ is the Dingle damping factor (with $T_D$ the Dingle temperature), and $R_S = \cos\left(\pi g \frac{m^*}{2m_0}\right)$ is the spin reduction factor ($m^*$ is the effective mass of electron and $g$ is the *Landé* factor). The exponent $\lambda$ is 0 for a two-dimensional (2D) Fermi surface (FS) and 1/2 for a three-dimensional (3D) FS [42]. In addition, $\Phi = \frac{\Phi_B}{2\pi} + \delta$, where $\Phi_B$ is the Berry phase and $\delta = 0$ for a 2D and $\pm 1/8$ for a 3D FS (+/- sign corresponds to the minima (+)/maxima (-) of the cross-sectional area of the FS for the case of an electron band; for a 3D hole band, the sign of δ is opposite) [43].

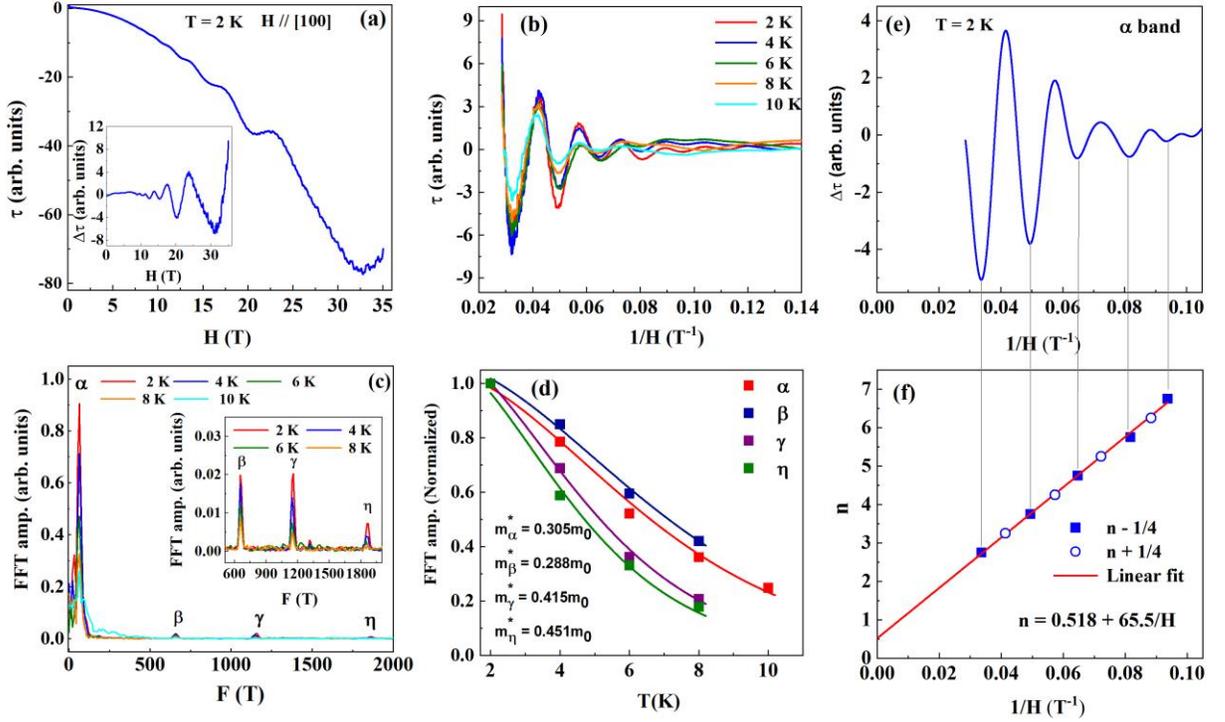

**FIG. 3. de Haas-van Alphen oscillations in magnetic torque of PdTe.** (a) Field dependence of the magnetic torque of PdTe at 2 K under $H // a$. Inset: $\Delta\tau$ vs. $H$. (b) $\Delta\tau$ plotted as a function of $H^{-1}$ at the indicated temperatures. (c) Fast Fourier Transformation (FFT) of the oscillatory torque presented in (b). The inset shows the enlarged FFT amplitude for the β, γ and η bands. (d) Temperature dependence of the FFT amplitudes for respective bands as indicated. Solid lines are the fit with thermal damping term of the LK formula. (e) de Haas-van Alphen oscillation at 2 K after applying a low pass filter of 100 T. (f) Landau fan diagram constructed from dHvA oscillation at 2 K for the α band.

As displayed in Fig. 3(c), the FFT amplitude for each oscillation decreases with increasing temperature. While the FFT spectrum is dominated by the low frequency (α band), higher frequencies can also be well resolved up to 8 K as shown in the inset of Fig. 3(c). Figure 3(d) displays the temperature dependence of the normalized FFT amplitude for four frequencies. From a fit of the thermal damping factor $R_T$ (see Eq. (1)) to the temperature dependence of the FFT amplitude, we obtain $m_\alpha^* = 0.305 m_0$, $m_\beta^* = 0.288 m_0$, $m_\gamma^* = 0.415 m_0$, $m_\eta^* = 0.451 m_0$. The effective mass for α, β and γ is consistent with the calculated value but smaller for the η band (see Table I). Using the Onsager relation, $F = (\hbar/2\pi e)S$, we estimate the extremal cross-section area of the Fermi surface ($S$) normal to the magnetic field direction for each frequency. Assuming the circular cross-section, the respective Fermi wave vector ($k_F$) is estimated. Correspondingly, the Fermi velocity $v_F = \hbar k_F/m^*$ is also calculated for each band, which is listed in Table I.

Analysis of the field dependence of the oscillation amplitude at a given temperature can provide information about the Dingle temperature $T_D$ through the Dingle plots of dHvA oscillations (see Fig. S3, Supplementary Material). From $T_D$, the quantum relaxation time, $\tau_q$, can be estimated through the relation $\tau_q = \frac{\hbar}{2\pi T_D k_B}$, which is proportional to the quantum mobility $\mu_q = \frac{e\tau_q}{m^*}$ [44]. The calculated results are listed in Table I for each band. Among four bands, the α band has the largest quantum mobility $\mu_q$ owing to its low $T_D$. On the other hand, the Fermi velocity of the η band is largest, which is attributed to its large $k_F$.

Table I: Parameters obtained from the dHvA oscillations for $H \mathbin{/\mkern-5mu/} a$ including the oscillation frequency ($F$), the Fermi wave vector ($k_F$), the effective mass ($m^*$), the Fermi velocity ($v_F$), the Dingle temperature ($T_D$), the quantum relaxation time ($\tau_q$), and the quantum mobility ($\mu_q$). The calculated dHvA frequency ($F$), effective mass ($m^*$) and carrier type (electron(e)/hole(h)) are also listed.

| Band | Experiment ($H \mathbin{/\mkern-5mu/} a$) | | | | | | | | Theory ($H \mathbin{/\mkern-5mu/} a$) | | |
|---|---|---|---|---|---|---|---|---|---|---|---|
| | $F$(T) | $k_F$ (Å$^{-1}$) | $m^*/m_0$ | $v_F$ ($10^5$ m/s) | $T_D$ (K) | $\tau_q$ ($10^{-13}$ s) | $\mu_q$ (m$^2$V$^{-1}$s$^{-1}$) | $\Phi_B$ (π) | $F$(T) | $m^*/m_0$ | e/h |
| α | 65.62 | 0.044 | 0.30(5) | 1.67 | 12.19 | 0.99 | 0.058 | 0.80 | 64.7 | 0.2544 | e |
| β | 658.4 | 0.141 | 0.28(8) | 5.66 | 23.88 | 0.50 | 0.030 | 0.25 | 671.2 | 0.2124 | e |
| γ | 1154 | 0.187 | 0.41(5) | 5.20 | 19.46 | 0.62 | 0.026 | 0.90 | 1141.9 | 0.3083 | h |
| η | 1867 | 0.238 | 0.45(1) | 6.08 | 18.39 | 0.65 | 0.025 | 0.40 | 2631.4 | 0.5409 | h |

The phase analysis of these dHvA oscillations can reveal the topological properties of the associated carriers. For such analyses, we isolate Δτ for each frequency via the filtering process [45] and determine the Berry phase of the carriers in each band. Figure 3(e) shows Δτ versus 1/$H$ for the α band. The Landau level fan diagram is then constructed, in Fig. 3(f), by assigning the oscillation minima to $n$-1/4 and maxima to $n$+1/4, where $n$ is the Landau level index [40]. As shown in Fig. 3(f), $n(H^{-1})$ can be described with the Lifshitz-Onsager quantization criterion [40, 43] with $n = 0.518 + 65.5/H$. The intercept of the linear equation gives the Berry phase as ($\Phi_B^\alpha/2\pi$) + δ = 0.518 while slope gives the frequency $F_\alpha$ = 65.5 T which is in excellent agreement with that obtained from the FFT spectra. According to Fig. 2(c), $F_\alpha$ corresponds to the minimum of the α band Fermi surface for $H \mathbin{/\mkern-5mu/} a$. With this and its 3D electron-type nature, we set δ = 1/8 and obtain the Berry phase $\Phi_B^\alpha = 0.78\pi$, a topologically nontrivial phase. With the similar manner, the Landau fan diagram constructed for the β band gives frequency $F_\beta$ = 667.3 T and $(\Phi_B^\beta/2\pi) + \delta$ = 0.743 (see Fig 5(a-b)). Since the β band is 3D electron type with a maximum FS for *H // a*, $\delta$ = -1/8, giving raise to the Berry phase $\Phi_B^\beta$ = 1.75π. This value is close to 2π, likely reflecting trivial topology. In view of the band structure shown in Fig. 2(b), the β band is not involved in the linear crossing, even though it meets with the α band at the A point and has a close proximity to α band at the K point in the Brillouin zone (see Fig. 2b). These close proximities between two electronlike bands in the high symmetry point can promote cross-pairing (formation of Cooper pairs with electrons originating from α and β bands) as illustrated in MgB$_2$ and Ba$_{0.6}$K$_{0.4}$Fe$_2$As$_2$ [46]. At present, existing experimental results including the ARPES work [26] can not unambigiously determine which band(s) contribute the observed superconducting properties, thus required further theoretical investigation as discussed in Refs. [47, 48]. Nevertheless, the multiple Fermi-surface packets with non-trivial topology we found here exclude conventional single-band isotropic *s*-wave superconductivity scenario [49, 50] for PdTe.

The Landau fan diagrams constructed for the γ and η bands are presented in Figs. S4 ((a-b), Supplementary Material). Comparing the linear fits with the Lifshitz-Onsager quantization criterion, we obtain $F_\gamma$ = 1151.5 T and $(\Phi_B^\gamma/2\pi) + \delta$ = 0.719 for the γ band, and $F_\eta$ = 1869.6 T and $(\Phi_B^\eta/2\pi) + \delta$ = -0.07 for the η band. These frequencies are in good agreement with that obtained from FFT analysis (see Table I). According to band structure calculations, both the γ and η bands are 3D hole type. When *H // a*, there is FS maximum for the γ band. We thus set $\delta$ = 1/8, which leads to Berry phase $\Phi_B^\gamma$ = 1.18π, a non-trivial Berry phase. The nontrivial topology of the γ band is consistent with our band calculations, which shows linear crossing with the η band along the Γ-A direction (Fig. 2(d)). This crossing is protected under the $C_{3z}$ rotational symmetry. For the η band there is a maximum for the smaller pocket centered at the A point for *H // a*. We thus assign $\delta$ = 1/8, which gives Berry phase $\Phi_B^\eta$ = -0.4π, suggesting that the small Fermi surface pocket is trivial.

**Fermi surface topology**

To get more insight into the Fermi surface topology of PdTe, we perform the angle dependence of the dHvA oscillations. Figure 4(a) shows the dHvA oscillations in Δτ of PdTe at *T* = 2 K at indicated angles. The angle θ is defined as depicted in the inset of Fig. 4(b). The dHvA oscillations can be observed in all angles between 0° and 90°, indicating the 3D character of the Fermi surface. With changing the direction of the field from $H // a$ to $H // c$, the dHvA oscillations change in both amplitude and peak position. Through the FFT analysis shown in the Fig. 4(b), we can track the spectrum with oscillations corresponding to the α, β and γ bands for all measured angles. The amplitudes associated with the large η band tends to vanish except for the extreme case i.e. at θ = 0° and 90°.

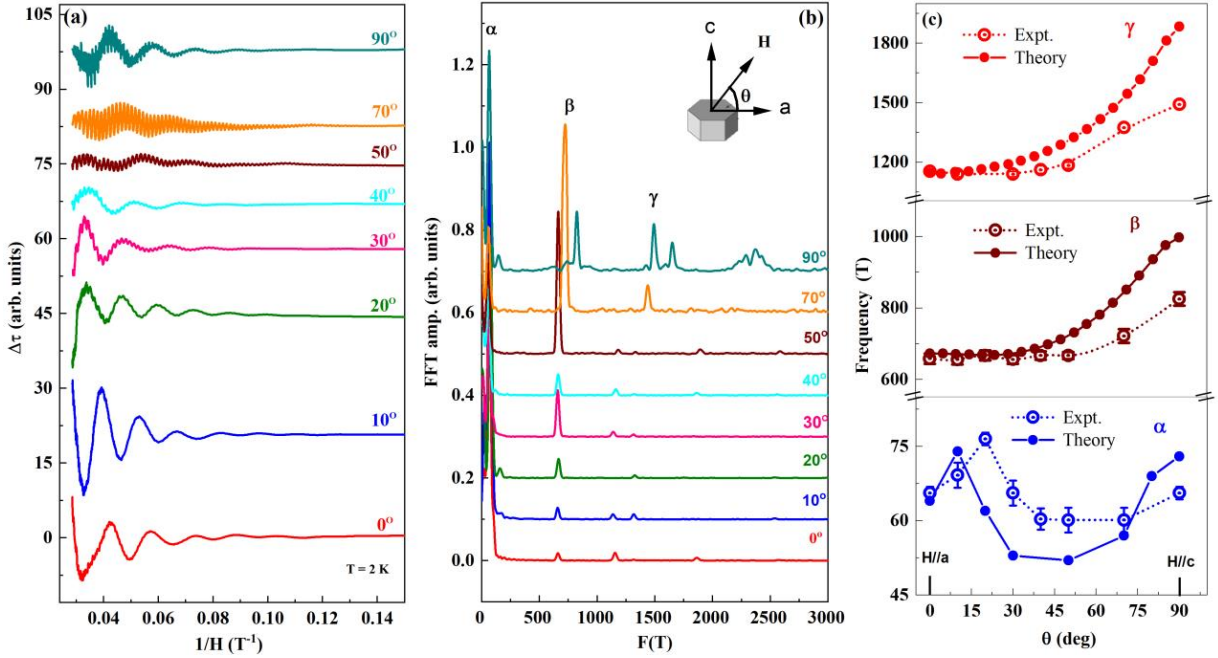

**FIG. 4. Angle dependence dHvA oscillations in PdTe.** (a) Magnetic torque of PdTe at $T = 2$ K after background subtraction plotted as $\Delta\tau$ $(H)$ versus $H^{-1}$ at indicated angles. A constant offset is added to the data for clarity. (b) The FFT spectra of the dHvA oscillations in (a). (c) The angle dependence of $F_\alpha$, $F_\beta$ and $F_\gamma$ from both experiment and calculations. The error bars are taken as the half-width at the half height of the FFT peaks.

Figure 4(c) displays the angular dependence of $F_\alpha$, $F_\beta$ and $F_\gamma$. $F_\alpha$ involves a nonmonotonic angle dependence as the field changes from $H // a$ to $H // c$, indicating the complex contour of the FS of this band, which is consistent with the calculated FS (Fig. 2(c)). The FFT amplitude varies monotonically with θ for for both the β and γ bands. The variation of the FFT amplitude is attributed to the spin reduction factor $\cos(\pi g\, m^*/2m_0)$ [40, 51], which includes the collective effect of the change in spin-orbit coupling strength as accounted by the $g$ factor and band curvature change as accounted by $m^*$ as described in Eq. (1). For comparison, the angle dependence of the $F_\alpha$, $F_\beta$ and $F_\gamma$ from DFT calculations is also plotted in Fig. 4(c). Note that the overall trend for three bands is similar. The enlarged discrepancy in high angles is likely due to the imperfect crystal alignment in experiment.

With data shown in Fig. 4, we can further examine the angle dependence of the Berry phase for the α, β and γ bands. While $\Phi_B^\alpha$ and $\Phi_B^\gamma$ have little change from $\theta = 0°$ to $90°$, $\Phi_B^\beta$ on the other hand varies dramatically. As shown in Fig. 5, $(\Phi_B^\beta/2\pi) + \delta$ change from 0.743 for $H \parallel a$ to 0.481 for $H \parallel c$. The latter corresponds to $\Phi_B^\beta \sim 1.2\pi$. This implies that there is likely a trivial to nontrivial topology crossover when rotating the field from the $a$ to $c$ direction. In recent ARPES measurements, the topological Fermi arc is observed at the (010) surface (i.e., the $ac$ plane) of PdTe [26]. At present, it is not obvious to determine how the β band is involved in topological properties.

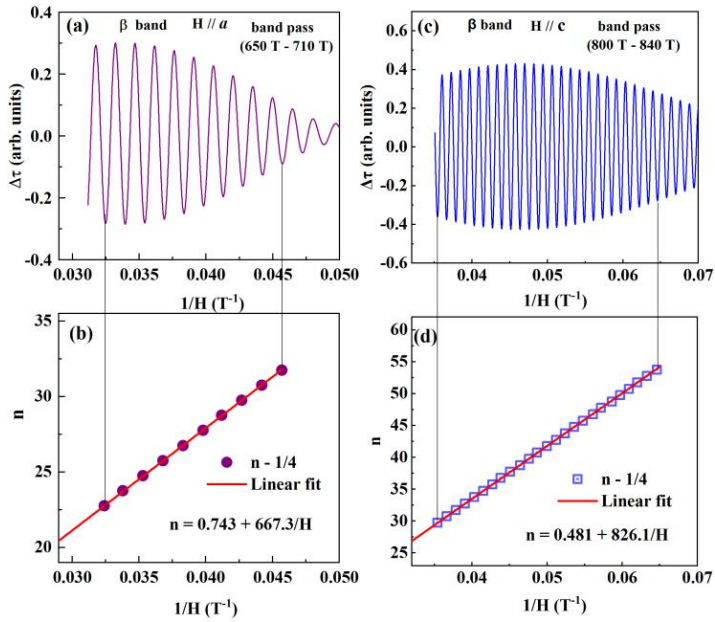

**FIG. 5. Landau fan diagrams.** Landau fan diagrams for the β band in the conditions of $H \parallel a$ (a-b) and $H \parallel c$ (c-d) constructed from the dHvA oscillations at $T = 2$ K. The numbers in the parenthesis in frame (a) and (c) denote the range of band pass filter used.

To further identify the topological nature of PdTe, we calculate the parity eigenvalues of $\alpha, \beta, \gamma, \eta$, and the surface spectral weight throughout the (100) surface Brillouin zone using the semi-infinite Green's function approach (Fig. S6, Supplementary Material, [52]). We notice that $\alpha$ and $\beta$ carry opposite parity eigenvalues to $\gamma$ and $\eta$, which imply the existence of the topological surface states between them based on the topological band theory. Our surface state calculations show a topological Dirac surface state in the bulk gap that forming α and β bands around the Γ point (Fig. S6(d)(g), Supplementary Material). In addition, we observe the surface states that emerge out of the Dirac node that forming η and γ bands, suggesting the nontrivial topology of this Dirac state (Fig. S6(e)(f), Supplementary Material). The surface state caclualtions are consistent with the parity eigenvalue analysis and support the Berry phase measurements from our experiments.

## Conclusions

In summary, we have investigated the physical properties of superconducting PdTe with $T_c \sim 4.25$ K in both the normal and superconducting states via magnetic torque and specific heat measurements, and first principles calculations. Below $T_c$, the electronic specific heat initially decreases in $T^3$ behavior (1.5 K < $T$ < $T_c$), consistent with the scenario that the superconducting gap consists of nodes as identified by ARPES [26]. However, the deviation of the electronic specific heat from the $T^3$ dependence below 1.5 K requires us to seek for alternative explanation. Using the two-band model, the superconducting specific heat can be well described with two energy gaps $\Delta_1 = 0.372$ meV and $\Delta_2 = 1.93$ meV with the larger $\Delta_2$ and specific heat jump $\Delta C_{es}$ indicating towards the strong electron-phonon coupling limit. The calculated bulk band structure consists of two electron bands (α and β) and two hole bands (γ and η) at the Fermi level. Detailed analysis of the dHvA oscillations observed in magnetic torque allows us to identify these four bands. By constructing the Landau fan diagram for each band, we extract the Berry phase, which is nontrivial for the α and γ bands, but a crossover for the β band from trivial at $H // a$ to nontrivial at $H // c$. Although further investigation is necessary to distinguish surface and bulk properties, the current investigation and recent ARPES work [26] strongly suggest that PdTe is a candidate for unconventional superconductivity including (1) nontrivial topology, (2) two-band scenario, and (3) superconducting gap with nodes in bulk but nodeless on surface.

## Methods

**Sample synthesis and structural characterization:** Single crystals of PdTe were grown using the method similar to that described in Refs. [26, 27]. The starting material, Pd powder (99.95%, Alfa Aesar) and Te powder (99.99%, Alfa Aesar) was mixed together in a ratio of Pd : Te = 1 : 1 and placed into an alumina crucible, which was then sealed in a quartz tube under vacuum. The whole assembly was heated to 1000 °C at a rate of 60 °C/h in a furnace, held at 1000 °C for 72 hours. The temperature was then lowered to 650 °C at a rate of 2 °C/h, and the furnace was turned off allowing to cool down to room temperature. Single crystals with typical size ~ $1.5 \times 1 \times 0.5$ mm$^3$ were obtained (shown in the inset of Fig. S2 (b), Supplementary Material). The structure of as-grown crystals was examined through powder (crushed single crystals) x-ray diffraction (XRD) measurements using a *PANalytical* Empyrean x-ray diffractometer (Cu K$_\alpha$ radiation; $\lambda$ = 1.54056 Å). All the diffraction peaks can be indexed under the NiAs-type hexagonal structure (space group $P6_3/mmc$) with the lattice parameter $a = b = 4.152(2)$ Å and $c = 5.671(2)$ Å (Supplementary Material, Fig. S2(a), consistent with the previously reported values [27].

**Electrical resistivity and specific heat measurement:** The electrical resistivity and specific heat were measured in a Physical Properties Measurement System (PPMS-14 T, Quantum Design) with a dilution refrigerator insert capable of cooling down to 50 mK. The electrical resistivity was measured using the standard four-probe technique. Thin platinum wires were attached to the single crystal sample using a silver epoxy (Epotek H20E). An electric current of 1 mA was used for the transport measurements.

**Magnetization and magnetic torque measurement:** The magnetization measurements were carried out in a magnetic property measurement system (MPMS-7 T, Quantum Design). Magnetic torque measurements were performed using the piezotorque magnetometry with a field up to 35 T at the National High Magnetic Field Laboratory (NHMFL) in Tallahassee, Florida, USA. The samples were mounted on self-sensing cantilevers and the cantilevers were placed in a $^3$He cryostat. Piezotorque magnetometry was performed with a balanced Wheatstone bridge that uses two piezoresistive paths on the cantilever as well as two resistors at room temperature that can be adjusted to balance the circuit. The voltage across the Wheatstone bridge was measured using a lock-in amplifier (Stanford Research Systems, SR860).

**First-principle calculations:** The electronic band structure of PdTe was computed using the projector augmented wave method [53] as implemented in the VASP package [54] within the generalized gradient approximation (GGA) schemes [55]. Experimental lattice parameters were used. A $21 \times 21 \times 21$ Monkhorst Pack $k$-point mesh was used in computations with a cutoff energy of 400 eV. The spin-orbit coupling (SOC) effects were included self-consistently. To compute the Fermi surface, energy bands were interpolated by mapping the electronic states onto a set of Wannier functions [56] using VASP2WANNIER90 interface [57]. We use Pd $d$-orbital and Te *p*-orbital to construct Wannier functions without performing the procedure for maximizing localization. The dHvA frequencies and effective masses were calculated by the SKEAF code [58] with the Fermi surface information.

## Data availabity

All data that support the findings of this study are available from the corresponding author upon reasonable request.

## Acknowdelgements

This material is based upon work supported by the US Department of Energy under EPSCoR grant DE-SC0012432 with additional support from the Louisiana Board of Regents. T.-R.C. was supported by the 2030 Cross-Generation Young Scholars Program from the National Science and Technology Council (NSTC) in Taiwan (program no. MOST111-2628-M-006-003-MY3), National Cheng Kung University (NCKU), Taiwan, and the National Center for Theoretical Sciences, Taiwan. This research was supported, in part, by Higher Education Sprout Project, Ministry of Education to the Headquarters of University Advancement at NCKU. A portion of this work was performed at the National High Magnetic Field Laboratory, which is supported by National Science Foundation Cooperative Agreement No. DMR-1644779 and the State of Florida.

## Author contributions

R. J. designed research; R. C. synthesized the sample and conducted physical property measurements with assistance from R. J., L. X, D. E. G and A. B. K., and P. V. S. R and T.-R. C performed the first-principle calculations. R. C. and R. J. wrote the manuscript with contributions from all the authors.

## Competing interests

The authors declare no competing interests.

## Additional information

Supplementary material contains structure details, Dingle plots of α, β, γ, and η bands, Landau fan diagrams for γ and η band. Orbital projected band structures and surface band structures.

**Correspondence** and requests for the materials should be addressed to R. J.

## Supplementary Material
**Structure details:**

PdTe crystalizes in the NiAs-type hexagonal Bravais lattice with lattice parameters $a = b = 4.1522$ Å and $c = 5.6712$ Å with the space group $P6_3/mmc$ (No:194), as shown in Fig.S1(a). In this structure, each Pd atom is surrounded by six Te atoms and forms tilted octahedral local structures as shown in Fig.S1(b). The structure contains two formula units with four atoms in the unit cell [1].

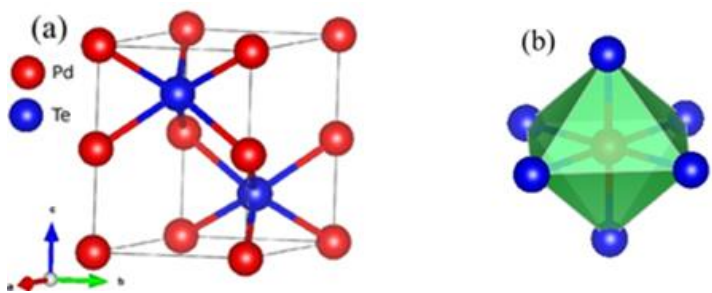

FIG. S1 (a) The crystal structure of PdTe. The red and blue spheres represent the Pd and Te atoms, respectively. (b) Octahedral local structure of PdTe.

As grown single crystals [1-3] were screened for their orientiatons via x-ray diffraction (XRD) measurements in a PANalytical Empyrean x-ray diffractometer (Cu K$_\alpha$ radiation; $\lambda$ = 1.54056 Å). The XRD pattern from a (00$l$) surface orientation is shown in Fig. S2(b). Few single crystals were crushed into fine powder and powder XRD measurement was performed. Fig. S2(a) shows the powder XRD pattern where all the diffraction peaks can be indexed under the NiAs-type hexagonal structure (space group $P6_3/mmc$) with the lattice parameter $a = b = 4.152(2)$ Å and $c = 5.671(2)$ Å consistent with the previously reported values [1].

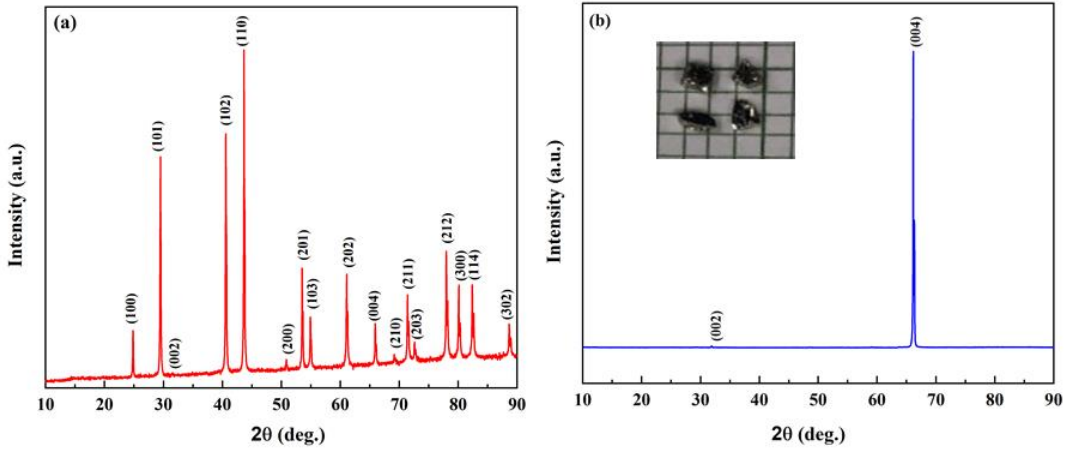

FIG. S2. (a) The powder X-ray diffraction (XRD) pattern of PdTe indexed in the NiAs-type hexagonal structure with space group P6$_3$/mmc (194). (b) The XRD pattern of a single crystal of PdTe where the indexed peaks are from (*0 0 l*) plane. Inset: typical PdTe single crystals.

**Dingle plots:**

To get information about the dynamics of the carries, Dingle temperature ($T_D$) is estimated through the Dingle plots. Figs. S3(a-d) present the Digle plots constructed for corresponding bands from the dHvA oscillations at $T = 2$ K for the case of $H//a$. From the slope of the linear fit in the Dingle plot $T_D$ is obtained (displayed in the respective frame) which are listed on Table 1 in the main text.

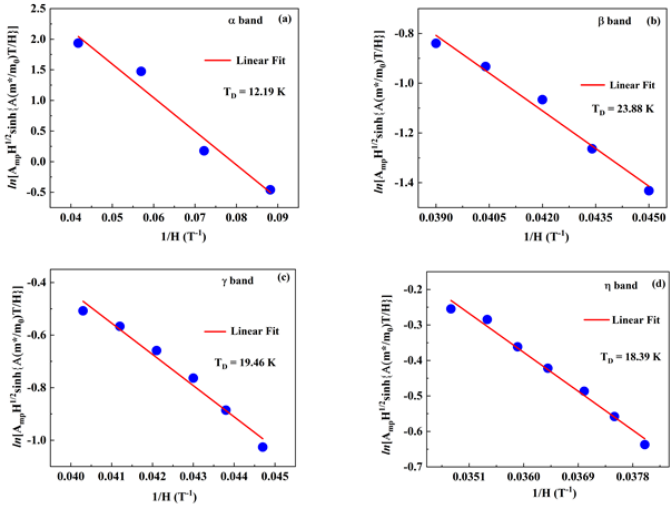

FIG. S3. The Dingle plots from dHvA oscillations in PdTe with $H//a$ at 2 K for α, β, γ and η bands.

**Landau fan diagrams for γ and η bands:**

The dHvA oscillations at $T = 2$ K under $H//a$ are used to construct the Landau fan diagrams. Suitable band-pass filtering [4] based on the frequency obtained via FFT analysis is applied to isolate the respective bands. Landau fan diagrams are then constructed by assigining the oscillation minima to $n-1/4$ and maxima to $n+1/4$, where $n$ is the Landau level index [5]. The Landau fan diagram thus obtained for the γ and η bands are presented in Figs. S4 (a-b). Comparing the linear fits with the Lifshitz-Onsager quantization criterion, corresponding frequency and Berry phase are obtained which are listed in Table 1 in the main text.

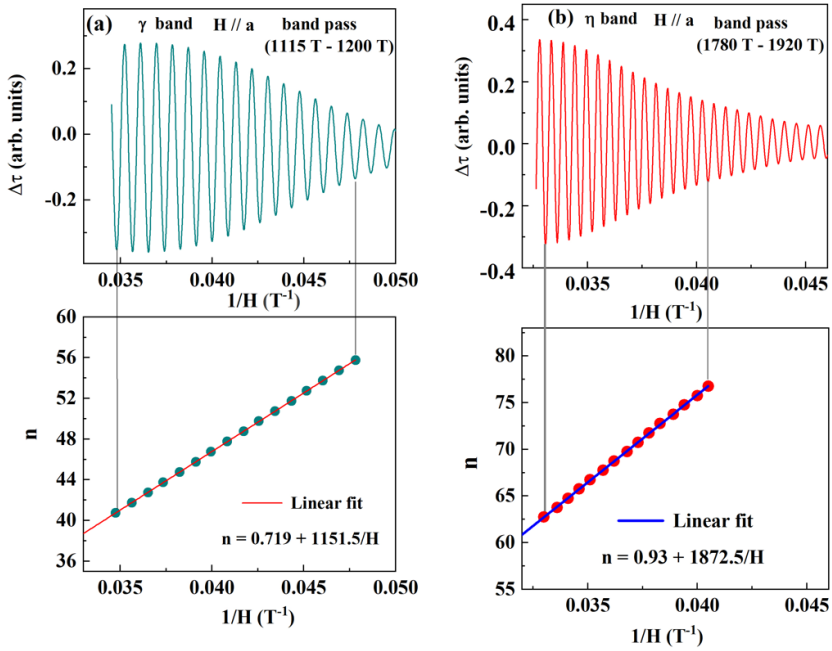

FIG. S4. Landau fan diagrams constructed from dHvA oscillation in magnetic torque of PdTe at 2 K under $H//a$ for (a) γ band and (b) η band.

**Orbital projected band structure:**

To identify the topological nature of PdTe, we calculate the surface spectral weight throughout the (100) surface Brillouin zone using the semi-infinite Green's function approach [6]. Fig. S5 shows the orbital projected band structure. We notice the topological Dirac surface state in the bulk gap forms α and β bands around the Γ point, indicating that the α and β bands carry opposite parity eigenvalue. In addition, we observe the surface states that emerge out of the Dirac node that forming η and γ bands, suggesting the nontrivial topology of this Dirac state.

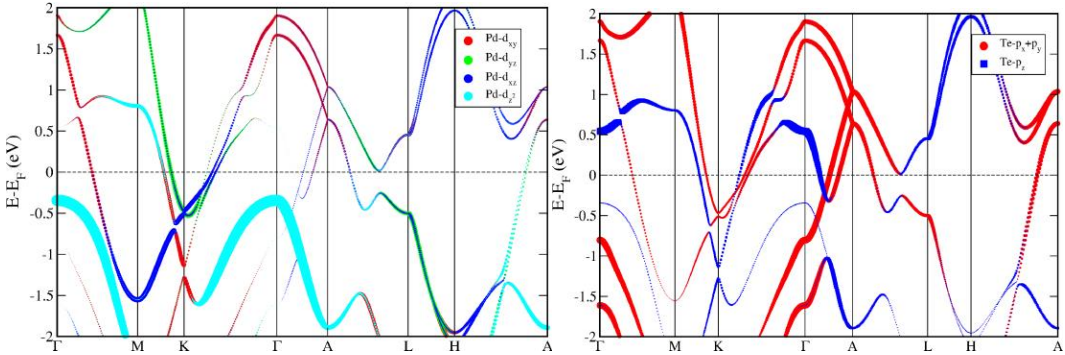

FIG. S5. Orbital projected band structure of PdTe with including spin orbit coupling.

**Surface band structures:**

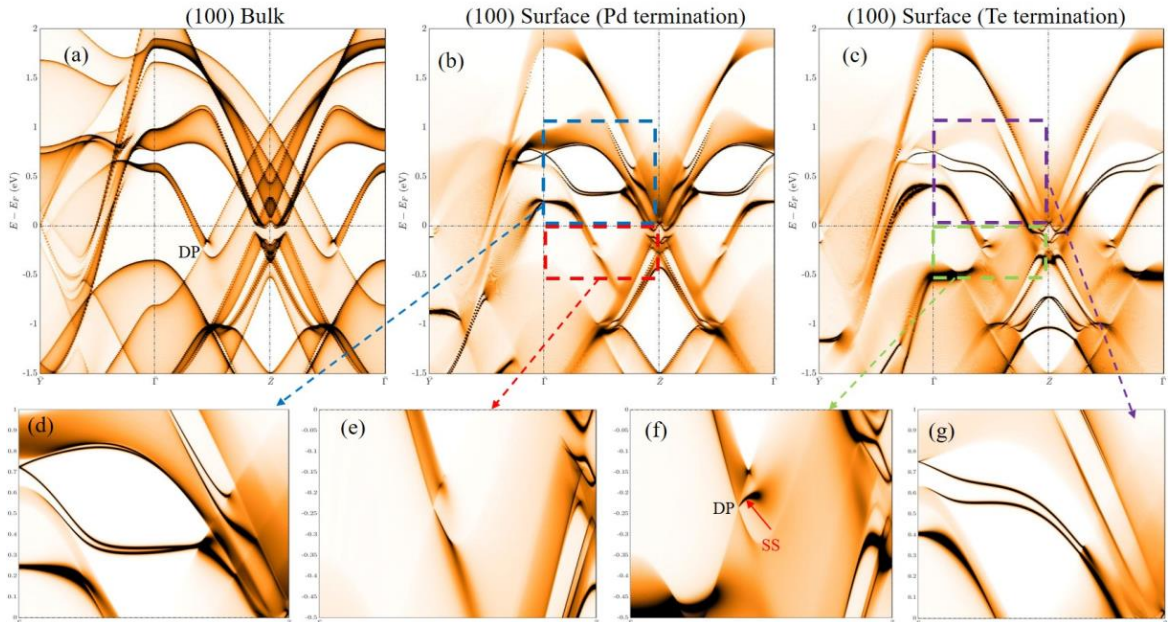

FIG. S6. Calculated (a) bulk, (b) Pd terminated surface and (c) Te terminated surface spectral weight for (100) surface of the BZ using semi-infinite Greens's function. (d) and (e) are zoomed areas as highlighted in (b). (f) and (g) are zoomed areas as highlighted in (c).

**References [SM]**
__________